\documentclass[preprint,10pt,times,3p]{elsarticle}

\usepackage{graphicx,multirow,nomencl,framed,xcolor,array,ragged2e}
\usepackage{amssymb}
\usepackage{amsmath}

\usepackage{lineno}
\usepackage{etoolbox}

\biboptions{sort&compress}

\journal{arXiv}
\makenomenclature
\begin{document}

\begin{frontmatter}

\title{Impact of liquid coolant subcooling on boiling heat transfer and dryout in heat-generating porous media}

\author[label1]{Aranyak Chakravarty\corref{cor1}}
\author[label2]{Koushik Ghosh}
\author[label2]{Swarnendu Sen}
\author[label2]{Achintya Mukhopadhyay}

\address[label1]{School of Nuclear Studies and Applications, Jadavpur University, Kolkata 700106, India}
\address[label2]{Department of Mechanical Engineering, Jadavpur University, Kolkata 700032, India}
\cortext[cor1]{Corresponding Author: \textit{E-mail}: aranyak.chakravarty@jadavpuruniversity.in}

\begin{abstract}
The present article discusses the impact of liquid coolant subcooling on multiphase fluid flow and boiling heat transfer in porous media with internal heat generation. Although extremely relevant and important, only limited studies are available in the open literature on the effects of coolant subcooling in heat-generating porous media and hence, this requires a detailed analysis. The analysis is carried out using a developed computational model of multiphase fluid flow through clear fluid and porous media considering the relevant heat transfer and mass transfer phenomena. Results suggest that the qualitative nature of boiling heat transfer from the heat-generating porous body, with subcooled coolants, remain similar to that observed with a saturated coolant. Quantitative differences are, however, observed as a result of solid-liquid convective heat transfer, and competing mechanisms of boiling and condensation heat transfer, when subcooled liquid coolant is present. It is observed that a substantially larger heating power density is required - as coolant subcooling is increased -  for achieving the same temperature rise in the porous body. Dryout power density at different coolant subcooling is also observed to follow a similar trend. The thermal energy removal limit is, hence, observed to be substantially enhanced in presence of subcooled coolants. Further, the dryout region within the porous body is observed to gradually shift towards its interior as the coolant subcooling is increased.    
\end{abstract}


\end{frontmatter}

\begin{table}[t]  
\begin{framed}
\nomenclature{$a_i$}{Interfacial area density (1/m)}
\nomenclature{$c_p$}{Specific heat capacity (J/kg.K)}
\nomenclature{$F$}{Solid-fluid drag force (kg/m\textsuperscript{2}.s\textsuperscript{2})}
\nomenclature{$R$}{Interfacial momentum exchange coefficient (kg/m\textsuperscript{3}.s)}
\nomenclature{$g$}{Acceleration due to gravity (m/s\textsuperscript{2})}
\nomenclature{$h$}{Enthalpy (J/kg)}
\nomenclature{$\alpha$}{Volume fraction}
\nomenclature{$\rho$}{Density (kg/m\textsuperscript{3})}
\nomenclature{$k$}{Thermal conductivity (W/m.K)}
\nomenclature{$\varepsilon_f$}{Porosity}
\nomenclature{$\mu$}{Viscosity (kg/m.s)}
\nomenclature{$Q'''$}{Volumetric heat transfer rate (W/m\textsuperscript{3})}
\nomenclature{$V$}{Velocity (m/s)}
\nomenclature{$T$}{Temperature (K)}
\nomenclature{$p$}{Pressure (Pa)}
\printnomenclature
\end{framed}
\end{table}

\section{Introduction}
\label{sec:introduction}
Any material comprising a solid matrix, along with interconnected voids (or pores), is characterised as porous media. Some common examples of porous media include soil, sand, coal, skin/membranes/cells in living beings, industrial filters, concrete etc. Such widespread and varied occurrence in natural as well as industrial scenarios have naturally led to extensive research on porous media, with major thrust on characterising various aspects of transport in porous media. A particular focus area has been on improving heat removal from different systems involving porous media and having inherent heat generation and/or extremely large heat fluxes. The present analysis deals with the thermal characterisation of one such system involving heat-generating porous media.

Fundamental investigations into transport characteristics of single-phase flow in systems involving porous media with internal heat generation have observed the existence of buoyancy-driven flow leading to the establishment of natural and mixed convective flow situations \cite{haajizadeh,baytas,ejlali,prasad,du,chakravarty2016,chakravarty2017,chakravarty2018,chakravarty2019,chakravarty2020}. The transport characteristics have been observed to depend on the heat source and permeability of the medium. System geometry, thermal-physical properties of the materials and existence of external flow have also been observed to significantly influence the transport characteristics.

In contrast to the voluminous research available on single phase flows, only a handful of studies have attempted to fundamentally study multiphase transport phenomena in heat-generating porous media. However, several fundamental analyses are available on multiphase flow in porous media without consideration of heat generation \cite{cornwell,fukusako,li2011,zhou,daurelle,waite,yeo,benard,ramesh}. It is evident from these investigations that the boiling heat transfer characteristics in porous media approach that of clear fluids as the particle sizes are progressively made larger. However, temperature rise in the porous media remained much lower when compared to the temperature rise in clear fluids, under identical heat flux conditions. The critical heat flux (CHF) in a porous structure is, as such, encountered at a lower temperature as compared to clear fluids. 

It is, however, expected that internal heat generation will substantially influence the nature of boiling heat transfer and associated transport phenomena. Thus, it becomes critically important to fundamentally characterise the transport process for multiphase flow in porous media with heat generation.

\begin{figure*}
\centering
	\includegraphics[scale=0.7]{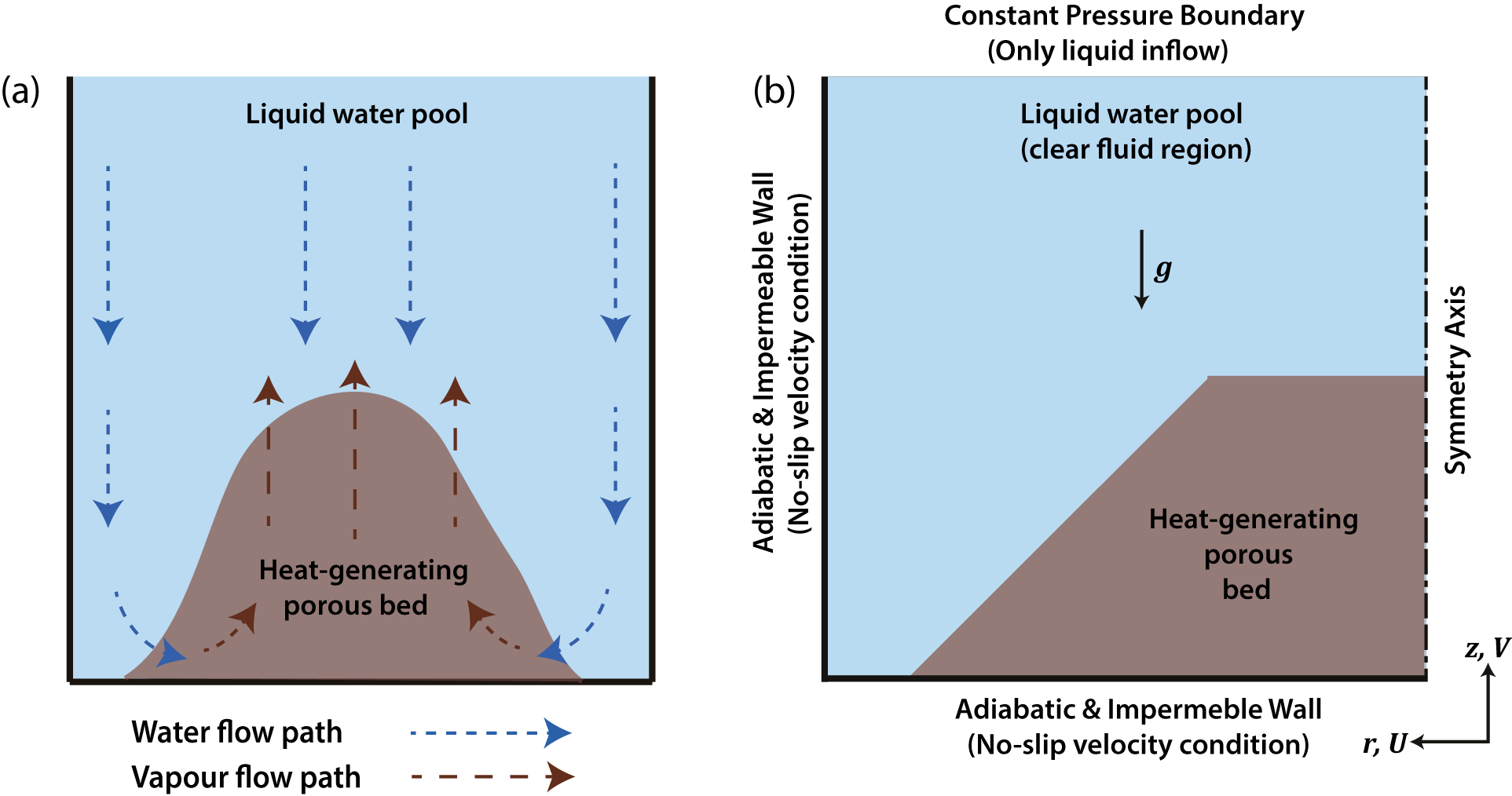}
	\caption{(a) Schematic representation of multiphase fluid flow in a pool of liquid water containing a heap-like porous bed with internal heat generation. Arrows denote the typical flow directions of liquid water and water vapour in the pool and the porous bed. (b) Approximate structure of the heap-like porous bed used in the present study. Symmetry condition allows computation using only half the actual domain. Relevant dimensions: Height \& width of cavity = 0.3 m, Bed height = 0.15 m, Bed bottom radius = 0.25 m, Bed top radius = 0.1 m, Slant angle = $45^{\circ}$, Bed volume = 0.01532 m$^3$.}
	\label{f:schematic}
\end{figure*}

Majority of investigations on multiphase flows in systems involving heat-generating porous media primarily deal with determination of the critical heat flux (CHF) and/or the dryout heat flux (DHF) in different situations \cite{raverdy,huang,rahman,bachrata,burger,schafer,lindholm,taherzadeh,yakush,miscevic}. WHile CHF quantifies the maximum thermal energy that can be removed from a system, DHF quantifies the minimum thermal energy level at which a system becomes uncoolable. These quantities are, therefore, used as indicators for determining whether adequate thermal energy can be removed from the porous medium. The major application of these indicators are in nuclear reactor safety - assessing heat removal from porous debris with decay heat generation, that are formed due to core meltdown accidents. Such assessments are extremely important due to the potential hazards that may arise out of failure to contain the progression of the accidents. In this regard, extensive research have identified CHF/DHF in different situations and the effect of various relevant parameters (for e.g. operating pressure, particle size, porosity, debris structure, coolant flooding mechanism etc.) on CHF/DHF. Surprisingly, only sparse information can be found in the public domain regarding the impact of coolant subcooling on CHF/DHF \cite{takasuo_rep}, although it is expected to have a substantial influence on the heat removal mechanism.

As such, it becomes crucial to computationally assess thermal energy removal limitations from porous media with internal heat generation, under different situations. The complexity involved in numerical modelling of the various flow regimes, heat and mass transfer mechanisms associated with multiphase flow makes such assessments extremely challenging. Several specialised numerical codes have been developed over the years that focus on proper modelling of the involved transport processes. These codes have been utilised to assess the energy removal limitations for different situations. However, these analyses have generally not focused on detailed discussion of the underlying heat transfer mechanisms, which forms the basis of developing an understanding of the transport processes and future improvement of the numerical models. Experimental studies in this context are also not exhaustive.

Fundamental characterisation of multiphase transport phenomena in porous media with internal heat generation, thus, gains extreme relevance. Chakravarty et al. \cite{chakravarty2020multiphase,chakravarty2020pressure} fundamentally analysed the boiling heat transfer characteristics and the resulting multiphase transport process in a cylindrical cavity with a heat-generating porous bed through numerical simulations. The simulations were carried out using ANSYS FLUENT - a general computational platform - instead of using specialised codes. Results indicate that the boiling heat transfer characteristics remain qualitatively similar to that in a clear fluid medium. However, boiling was observed to be initiated in the porous media at much less superheats when compared to clear fluids and is an outcome of the abundance of nucleation sites in a porous structure. Dryout was observed to occur at much less power densities as compared to clear fluids. Heat transfer and dryout occurrence in the porous bed was also observed to be dependent on composition and structure of the porous media as well as on the system pressure and coolant subcooling. 

The present study is a continuation of the aforementioned analyses \cite{chakravarty2020multiphase,chakravarty2020pressure} and attempts to properly characterise the impact of coolant subcooling on the transport processes. As discussed, no proper analysis of the effect of liquid subcooling on boiling heat transfer in porous media with internal heat generation is available in the open literature, to the best of the author's knowledge. Investigations on the impact on subcooling in other situations, however, indicate that subcooling can substantially affect the heat transfer mechanism \cite{dhir1998,ono2007,el2016}. Analysing the impact of subcooling in heat-generating porous media, thus, becomes extremely relevant keeping in mind the application of such analyses in energy removal assessments.
 
Computations are carried out using a generalised computational model that can properly account for the complex nature of fluid transport, heat transfer as well as mass transfer in clear fluids and porous medium. The computational model is incorporated in ANSYS FLUENT using user-defined functions. The model is used for characterising boiling heat transfer and multiphase fluid transport phenomena in a realistic heap-shaped porous bed immersed in a pool of subcooled liquid coolant (see Fig. \ref{f:schematic}a). Internal heat generation is assumed within the porous bed. The characterisation is achieved using distributions of solid temperature and liquid volume fraction within the computational domain. A profile of temperature increase in the porous bed is also constructed with respect to heating power density for highlighting the impact of liquid subcooling. Dryout power density for different cases are identified using the profiles of temperature increase and transient behaviour of liquid volume fraction and temperature of the heat-generating solid phase. It is worthwhile to mention here that this computational model can also be utilised for investigating other phenomena having similar transport physics, with proper change in closure relations, wherever necessary.

\section{Problem Description}
\label{sec:problem_desc}

\subsection{Physical Configuration}
\label{sec:phys_conf}
The physical configuration considered in the present analysis is a situation hypothesised to be encountered during post-accident removal of decay heat from a porous debris bed in nuclear reactors. Figure \ref{f:schematic}b schematically represents the physical configuration being considered. A fluid-filled cylindrical cavity is assumed with adiabatic and impermeable walls at the side and bottom of the cavity. The top boundary of the cavity is subjected to a constant pressure condition which enables fluid transport into and out of the cavity. This is subjected to an additional condition that while both phases can flow out of the cavity, only the liquid coolant (water) can flow into the cavity. The direction of fluid transport is determined by the fluid pressure inside the cavity. 

The porous bed is considered to be situated at a central location on the bottom wall of the cavity and is considered to have a truncated conical shape, that approximates the typical heap-shaped structure of porous debris \cite{lin}. It is further assumed that the structure and the composition of the bed remains invariant with time. Dimensions of the porous bed and the cylindrical cavity are stated in Fig. \ref{f:schematic}. Relevant composition and material properties of the porous bed ($\rho_s=4200$ kg/m\textsuperscript{3}; $k_s=2.0$ W/m.K; $c_{p,s}=775$ J/kg.K) are assumed to be the same as in Takasuo et al. \cite{takasuo} . 

The entire configuration is further assumed to be symmetric about the cavity's central axis. Initially, it is assumed that the cavity is filled with subcooled liquid coolant (water) and all other phases are at the saturation temperature.

\subsection{Mathematical Formulation and Assumptions}
\label{sec:math_form}
It is easily understood from the physical configuration described in Section \ref{sec:phys_conf} that the overall transport process within the cylindrical cavity is a coupled phenomena that is dependent on the individual transport characteristics within the porous debris and the surrounding water pool. As such, the transport equations need to be solved for both these regions - with appropriate inter-regional coupling - for properly characterising the transport phenomena. 

Some simplifying assumptions are made while formulating the governing transport equations for either region, with proper care to ensure that the underlying physics of the phenomena do not get affected. The porous debris is assumed to have isotropic properties and a homogeneous composition. The solid phase of the porous bed is assumed to be comprised of perfectly spherical heat-generating particles. However, the solid phase is assumed not to undergo any phase change. The voids within the porous bed are assumed to remain saturated with the same fluid as in the surrounding fluid zone. Additionally, all constituent phases are assumed to have invariant thermo-physical properties, except density of the fluid phases. Fluid density is modelled following the Boussinesq approximation to account for the natural convective fluid motion arising out of the induced temperature differences.  

The governing transport equations considering the above stated assumptions are stated below. Equations \ref{eq:mass_fluid} – \ref{eq:en_fluid} and \ref{eq:mass_porous} – \ref{eq:en_f_porous} represent the transport equations for the fluid phases in the fluid zone and the porous bed, respectively. 

\flushleft{\textbf{\textit{Fluid zone}}}
\begin{equation}
\frac{\partial{(\alpha_j\rho_j)}}{\partial t} + {\nabla}\cdot{(\alpha_j\rho_j\vec{V}_j)} =\dot{m}_{kj}'''  ,
\label{eq:mass_fluid}
\end{equation}

\begin{equation}
\begin{aligned}
\frac{\partial{(\alpha_j\rho_j\vec{V}_j)}}{\partial t} +  {\nabla}\cdot {(\alpha_j\rho_j\vec{V}_j\vec{V}_j)} = -\nabla(\alpha_jp)+\mu_j\nabla^2\vec{V_j}+\dot{m}_{kj}'''\vec{V}_{kj}\\
+R_{kj}(\vec{V}_k-\vec{V}_j)+\alpha_j\rho_j\vec{g}-\nabla\cdot(\rho_j\vec{V}_j'\vec{V}_j'), 
 \label{eq:mom_fluid}
\end{aligned}
 \end{equation}

\begin{equation}
\frac{\partial{(\alpha_j\rho_j h_j)}}{\partial t} +  {\nabla}\cdot {(\alpha_j\rho_j\vec{V}_j h_j)} = \alpha_j k_j\nabla^2 T_j- Q_{ji}'''+ \dot{m}_{kj}''' h_{j,sat}-\nabla\cdot(\rho_jh_j'\vec{V}_j'), 
\label{eq:en_fluid}
\end{equation}

\flushleft{\textbf{\textit{Porous bed}}}
\begin{equation}
\frac{\partial{(\varepsilon_f\alpha_j\rho_j)}}{\partial t} +  {\nabla}\cdot {(\alpha_j\rho_j\vec{V}_j)} = \dot{m}_{kj}''',
\label{eq:mass_porous}
\end{equation}

\begin{equation}
\begin{aligned}
\frac{\partial{(\alpha_j\rho_j\vec{V}_j)}}{\partial t} +  {\nabla}\cdot \frac{(\alpha_j\rho_j\vec{V}_j\vec{V}_j)}{\varepsilon_f} = -\nabla(\varepsilon_f\alpha_jp)+\mu_j\nabla^2\vec{V_j}\\+\frac{1}{\varepsilon_f}(\dot{m}_{kj}'''\vec{V}_{kj}+R_{kj}(\vec{V}_k-\vec{V}_j))+\varepsilon_f\alpha_j\rho_j\vec{g}+\vec{F}_{sj} 
 \label{eq:mom_porous}
\end{aligned}
\end{equation}  
  
\begin{equation}
\frac{\partial{(\alpha_j\rho_j h_j)}}{\partial t} +  {\nabla}\cdot {(\alpha_j\rho_j\vec{V}_j h_j)} = \alpha_j\varepsilon_f k_j\nabla^2 T_j + Q_{sj}''' - Q_{ji}'''+ \dot{m}_{kj}''' h_{j,sat}, 
\label{eq:en_f_porous}
\end{equation}  

\justifying
Energy transport in the solid phase of the porous bed is modelled as - 
\begin{equation}
\frac{\partial{[(1-\varepsilon_f)\rho_s c_{p,s}T_s]}}{\partial t} =(1-\varepsilon_f)k_s\nabla^2 T_s+ Q_s''' - Q_{sl}''' - Q_{sv}''' - Q_{si}'''. 
\label{eq:en_s_porous}
\end{equation}

\justifying
Subscripts $j$ and $k$ in the aforementioned equations represent the primary and the secondary phase index, respectively - this may be liquid ($l$) or vapour ($v$) as the case may be. Subscript $s$ denotes the solid phase in the porous bed. Subscript $i$ denotes the liquid-vapour interface.

The term $R_{k,j}$ in the above equations denote the interfacial drag co-efficient between the fluid phases. The term $F_{s,j}$ in Eq. \ref{eq:mom_porous} takes into account the solid-fluid drag within the porous bed. $Q_s'''$ in Eq. \ref{eq:en_s_porous} denotes the rate of volumetric heat generation in the solid phase of the porous bed. $Q_{s,j}'''$ in Eq. \ref{eq:en_f_porous} denotes the convective heat transfer between the solid phase and the primary fluid phase, and is separately denoted as $Q_{s,l}'''$ (primary liquid phase) and $Q_{s,v}'''$ (primary vapour phase) in Eq. \ref{eq:en_s_porous}. $Q_{j,i}'''$ in Eqs. \ref{eq:en_f_porous} and \ref{eq:en_fluid}, and $Q_{s,i}'''$ in Eq. \ref{eq:en_s_porous} represents interfacial heat transfer from the primary fluid phase and boiling heat transfer, respectively.

\begin{figure}[!ht]
\centering
	\includegraphics[scale=0.6]{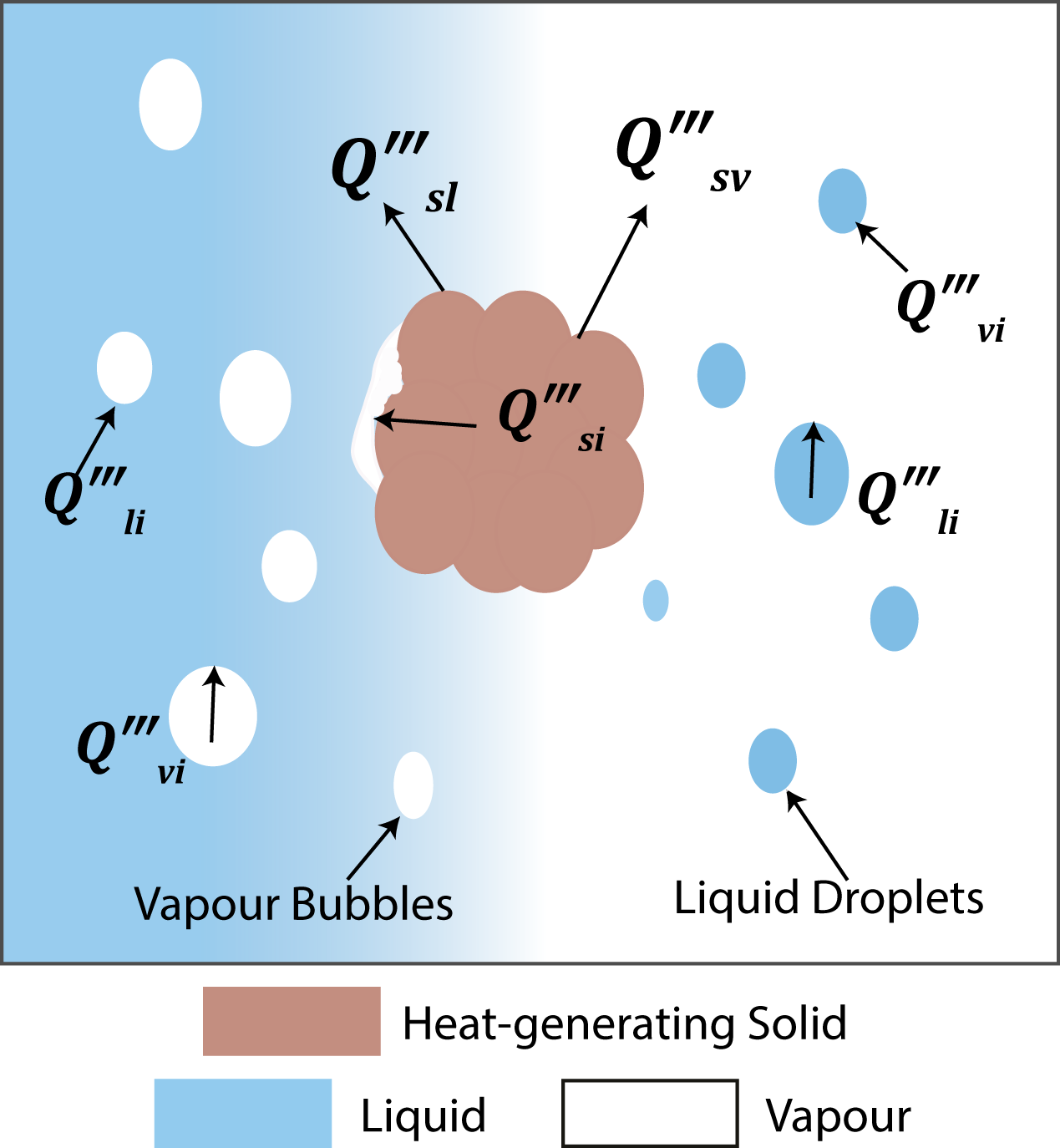}
	\caption{Schematic representation of different heat transfer mechanisms considered between the constituent phases.}
	\label{f:schematic_HT}
\end{figure}

Appropriate closure relations are utilised for modelling the different terms mentioned above subject to flow regime demarcation (see Table \ref{tab:flowregime}). The flow regime demarcation utilised in the present analysis is based on previous similar studies \cite{rahman,bachrata}. The correlations used are summarised in Table \ref{tab:correlations}. Figure \ref{f:schematic_HT} schematically shows the various heat transfer mechanisms considered in this analysis. Detailed discussion regarding the different correlations utilised and their implementation can be found in Chakravarty et al. \cite{chakravarty2020multiphase}. 

The term $\dot{m}_{kj}'''$ in the above equations represent the volumetric mass transfer rate between the fluid phases. It is estimated as

\begin{equation}
\dot{m}_{kj}'''=\frac{Q_{s,i}'''+Q_{l,i}'''+Q_{v,i}'''}{h_{v,sat}-h_{l,sat}}
\label{eq:mass_transfer}
\end{equation}

The effects of turbulence need to be accounted in the present problem in addition to the above discussed correlations and assumptions. Fluid flow remains laminar within the porous bed due to considerable resistance against fluid flow as a consequence of the small permeability of the porous bed. The same assumption is, however, not applicable for the clear fluid region. Hence, turbulence effects is considered only for the clear fluid region. Turbulence is modelled using the $k-\epsilon$ mixture turbulence model and integrated in Eqs. \ref{eq:mom_fluid} and \ref{eq:en_fluid}.

\begin{table}
\centering
\caption{Demarcation criteria for different regimes of fluid flow \cite{rahman,bachrata}}
\begin{tabular}{c c}
\hline
\textbf{Flow regime} & \textbf{Demarcation criteria}\\
\hline
Bubbly flow regime & $\alpha_v\leq 0.3$\\
\hline
Transition flow regime & $0.3<\alpha_v< 0.99$\\
\hline
Droplet flow regime & $\alpha_v\geq 0.99$\\
\hline
\end{tabular}
\label{tab:flowregime}
\end{table}

\begin{table*}
\centering
\caption{Correlations used for modelling different closure terms in the computational model}
\begin{tabular}{c m{4cm} m{4cm} m{5.2cm}}
\hline
\multicolumn{2}{c}{\textbf{Closure Term}} & \textbf{Correlation} & \textbf{Remarks}\\
\hline
\multirow{2}{*}{Interfacial drag} & Fluid zone ($R_{kj}$) & Schiller-Naumann \cite{schiller} & \multirow{2}{*}{Subjected to flow regime demarcation}\\

& Porous bed ($F_{sj},R_{kj}$) & Schulenberg-Muller \cite{schulenberg}  & \\
\hline
\multirow{3}{*}{Heat transfer} & Solid-fluid convective heat transfer ($Q_{s,j}'''$) & Ranz-Marshall \cite{ranz} & $\alpha_l\geq0.7$ (solid-liquid), $\alpha_v\geq0.99$ (solid-vapour)\\
& Boiling heat transfer ($Q_{s,i}'''$) & Rhosenow \cite{rohsenow}- Nucleate boiling; Bromley \cite{bromley} - Film boiling; Weighted average - Transition boiling & $T_s>T_{sat}$\\
& Liquid-vapour interfacial heat transfer ($Q_{j,i}'''$) & Ranz-Marshall \cite{ranz} & $0<\alpha_v<1$\\
\hline
\end{tabular}
\label{tab:correlations}
\end{table*}

\section{Numerical Procedure}
\label{sec:num_proc}

\subsection{Model implementation}
\label{sec:implementation}
The computational model discussed in Section \ref{sec:math_form} is implemented in ANSYS FLUENT. The governing equations (Eqs. 1 - 6) are solved using the Eulerian multiphase approach. Solid energy transport equation (Eq. 7) is solved as a separate equation in ANSYS FLUENT using user-defined functions and the solid phase temperature ($T_s$) as the unknown variable. The closure relations and related parameters are also implemented through user-defined functions and coupled to the relevant transport equations, as and when needed. 

\subsection{Identification of dryout condition}
\label{sec:dryout_identify}
The maximum amount of thermal energy that can be removed from any system is characterised by its CHF or DHF (see Section \ref{sec:introduction}). An important issue faced in computational assessment of this limit is to properly identify the CHF/DHF from numerical data. In the present analysis, this limit is identified by from the transient change of the minimum  and maximum magnitudes, respectively, of liquid fraction ($\alpha_{l,min}$) and temperature of the solid phase ($T_{s,max}$) within the porous bed. If $\alpha_{l,min}$ within the porous bed drops to zero and remains there for the subsequent time, and the corresponding $T_{s,max}$ shows a continued rise beyond the steady-state solid temperature reached at the immediately lower heat generation rate, it is inferred that dryout has taken place inside the porous bed. The corresponding volumetric heat generation rate is identified as dryout power density, while the immediately lower heat generation rate is termed as critical power density. The region of the bed where dryout has occurred is identified from the corresponding distributions $\alpha_l$ and $T_s$.

\subsection{Model validation}
\label{sec:validation}
The present analysis uses the same computational model as that reported in Chakravarty et al. \cite{chakravarty2020multiphase}. The model has been validated against reported experimental results for two important aspects - pressure drop during multiphase flow in porous media and dryout identification in heat-generating porous debris for different operating conditions (see Chakravarty et al. \cite{chakravarty2020multiphase} for details). Pressure drop validation ensures proper treatment of the flow regimes in porous media by the computational model. Dryout identification using the model ensures validation of the different heat transfer mechanisms assumed and the correlations utilised. However, no experimental or numerical studies were found which have studied situations with subcooled coolants in heat-generating porous beds. As such, no additional validation study could be carried out for situations with subcooled flow conditions. Nonetheless, the results obtained in the present study have been observed to remain qualitatively similar for boiling heat transfer in subcooled flow conditions in clear fluid medium (see Fig. \ref{f:dryout}b). 

\subsection{Solution algorithms and Grid-independency}
\label{sec:schemes}
Different numerical algorithms are utilised to solve various aspects of the problem. Pressure-velocity coupling is obtained with the phase-coupled SIMPLE algorithm. The transient terms of the transport equations are solved following the second-order implicit scheme, while the gradient terms are solved using a least squares cell-based method. The volume fraction calculation is obtained with the QUICK scheme. All other equations are solved with the second-order UPWIND scheme. 

A detailed grid-independence analysis has been conducted out before obtaining the results reported in this article. It has been determined from this analysis that a 3 mm nominal cell size and a time-step size of $10^{-3}$s accurately captures the transport mechanism and identifies dryout within the porous bed (see Chakravarty et al. \cite{chakravarty2020multiphase} for details). This time-step and grid configuration is, hence, used for carrying out the analyses reported in this article.

\begin{figure}[!ht]
\centering
	\includegraphics[scale=0.35]{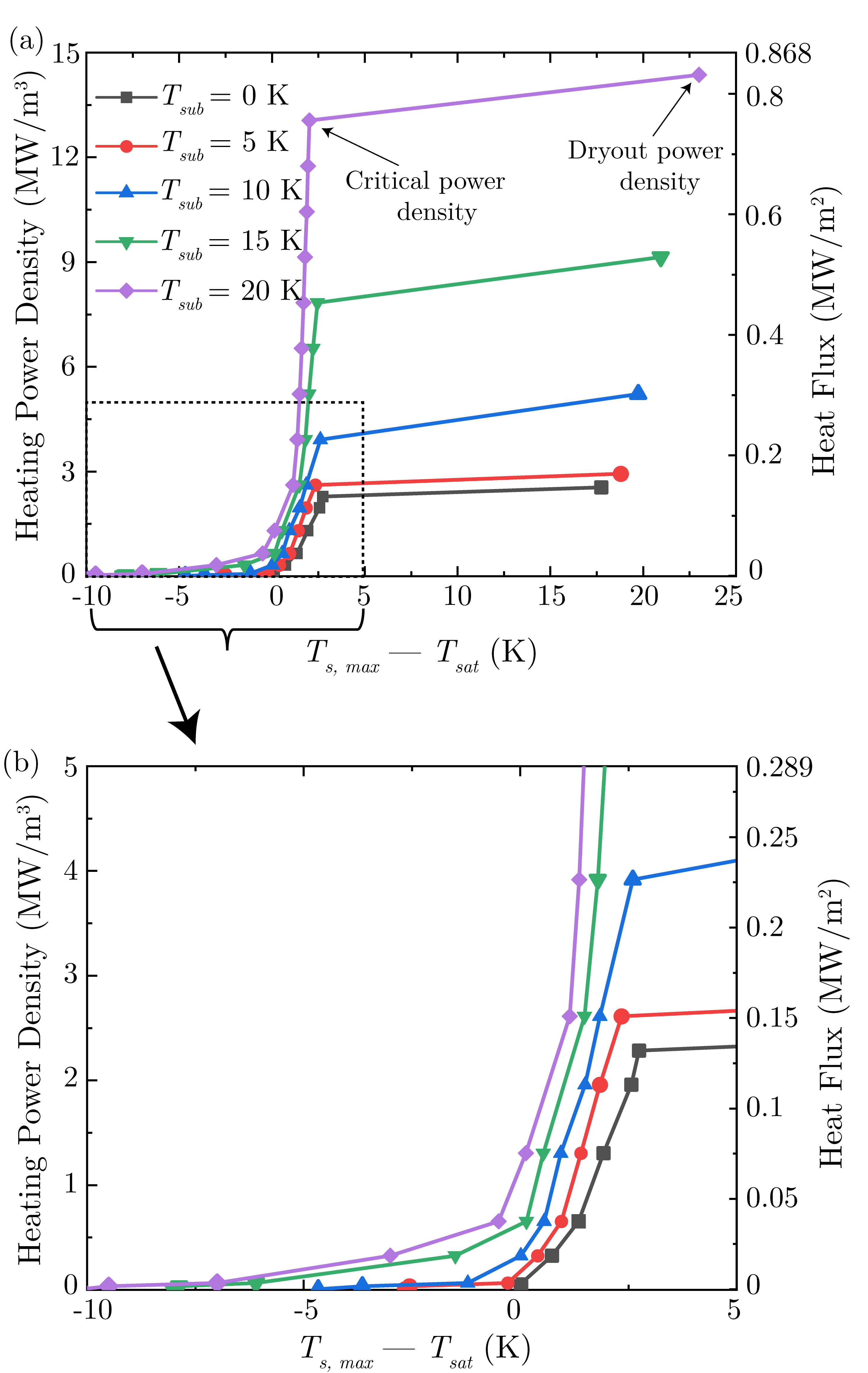}
	\caption{(a) Maximum rise in solid phase temperature within the porous bed - relative to the saturation temperature - as a function of heating power density and heat flux at different liquid subcooling. Total surface area of the porous bed in contact with the fluid zone is considered for heat flux calculation. A zoomed-in view of the plot is shown in (b) to adequately illustrate the change in temperature rise for different subcooled conditions.}
	\label{f:boilingcurve}
\end{figure}

\section{Results and Discussion}
\label{sec:results_discussion}

The effects of liquid coolant subcooling on heat removal from the heat-generating porous bed is assessed by considering the temperature of the in-flowing liquid coolant between the saturation temperature ($T_{sat}$) and a subcooling of 20 K i.e. ($T_{sat}-20$). System pressure for the entire analysis is maintained constant at 1.3 bar ($T_{sat} = 380.259K$). Average porosity and permeability of the porous bed are taken to be 0.39 and $9.59\times10^{-10}$ $\text{m}^2$ (similar to composition of porous debris found during post-accident situations in nuclear reactors), respectively, which remain invariant throughout the analysis. 

\subsection{Heat transfer characteristics}
\label{sec:HT_chars}

Figure \ref{f:boilingcurve}a plots the change in maximum temperature of the solid phase in the porous bed with respect to the saturation temperature i.e. ($T_{s,max} - T_{sat}$) with variation in heating power density for different coolant subcooling. It is evident from Fig. \ref{f:boilingcurve}a that an increase in heating power density results in a higher $T_{s,max}$ within the porous bed. The progressive rise in $T_{s,max}$ is, however, observed to have a non-linear nature that is dependent on the relative contribution of the involved heat transfer mechanisms. This is typical to the observed characteristics of boiling heat transfer in clear fluids \cite{dhir1998}. Although similar qualitatively, quantitative differences have been observed between the boiling curves in clear fluids and porous media, particularly in respect to the superheat required for onset of boiling. It has been experimentally observed that boiling starts at a much lower superheat in porous media \cite{fukusako}. Similar observations have been made in the present analysis as well. A detailed discussion of the boiling curve obtained using the present computational model and its similarity to boiling curves reconstructed from experimental data can be found in Chakravarty et al. \cite{chakravarty2020multiphase}. That discussion is, however, limited to the case of a saturated coolant only and hence, warrants further discussion on the effects of coolant subcooling.

It has been observed by Chakravarty et al. \cite{chakravarty2020multiphase} that the main parameter determining the mechanism of heat transfer in the porous bed, for saturated liquid coolant, is the inherent temperature increase inside the porous bed. The increase in temperature is an outcome of the volumetric heat generation in the solid phase of the porous bed. The temperature gradient created due to such temperature rise leads to convective heat transfer from the solid phase to the liquid coolant. Since the liquid coolant remains at $T_{sat}$, the transferred thermal energy from the solid phase leads to evaporative heat transfer from the liquid coolant causing vapour generation within the bed. Boiling heat transfer also plays an important role since the inherent temperature rise causes $T_s$ to increase beyond its initial value at $T_{sat}$. Convective heat transfer between the solid phase of the porous bed and vapour phase starts after vapour formation begins inside the porous bed. 

\begin{figure*}[!ht]
\centering
	\includegraphics[scale=0.4]{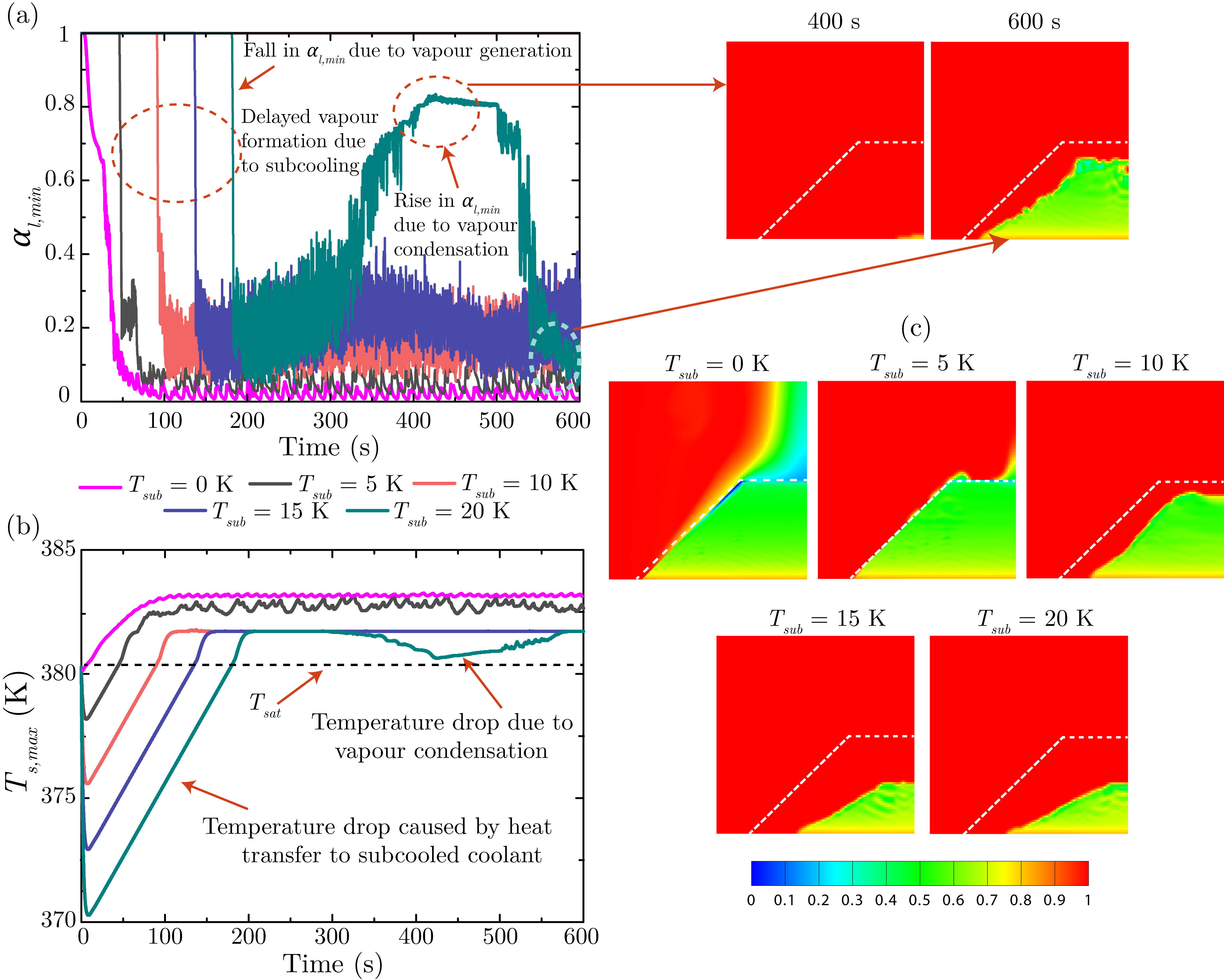}
	\caption{Transient variation in (a) $\alpha_{l,min}$ and (b) $T_{s,max}$, and (c) liquid volume fraction distribution for different coolant subcooling at a constant heating power density (2.4804 MW/m\textsuperscript{3}). It is observed that the steady-state temperature of the solid phase becomes gradually lower and the vapour content within the bed becomes increasingly larger with increase in coolant subcooling. This is primarily due to the effects of convective solid-liquid heat transfer and condensation heat transfer. Also, the vapour generated is observed to get progressively confined to the inner regions of the porous bed.}
	\label{f:dynamics}
\end{figure*}

The heat transfer mechanism is observed to change significantly when subcooled coolant is introduced into the domain. In addition to the heat generation induced temperature rise in the solid phase, the initial temperature difference - between the subcooled liquid coolant and the heat-generating solid phase - is observed to play an important role when subcooled coolant is considered. The existing temperature difference results in significant heat transfer from the solid phase to the subcooled liquid coolant. Thus, temperature of the liquid coolant increases while the solid phase temperature reduces, in spite of heat generation in the latter. This continues until a thermal equilibrium is reached between the solid and liquid phases. Thereafter, negligible convective heat transfer takes place between the liquid coolant and the solid phase and as such, the solid phase temperature again starts increasing due to the heat generation (see Fig. \ref{f:dynamics}b). 

As can be inferred from Fig. \ref{f:dynamics}b, the steady-state magnitude of $T_s$ is governed by the relative dominance of heat generation in the solid phase and coolant subcooling. Impact of heat generation remains small at low heating power densities and hence, the steady-state magnitude of $T_s$ remains below its initial value at $T_sat$ (see Fig. \ref{f:boilingcurve}). As the heat generation rate is gradually raised, the steady-state solid temperature progressively increases and exceeds $T_{sat}$ at a certain power density. Boiling heat transfer starts in such a situation leading to vapour generation within the porous bed. 

A closer look at Fig. \ref{f:boilingcurve} would reveal that the slope of the curve suddenly becomes steeper beyond a certain heating power density. This corresponds to the power density where boiling heat transfer starts to have a substantial contribution in the overall heat transfer. Nucleate boiling occurs resulting in substantial vapour generation within the porous bed. This leads to relatively less heat removal from the solid phase due to lower heat capacity of vapour as compared to liquids. The increase in steady-state magnitude of $T_s$ within the bed is, thus, observed to be relatively larger in the boiling heat transfer regime and results in the steeper temperature rise curve. Vapour generation continues to increase with further increases in the power density leading to progressively lower heat removal and as such, the temperature rise keeps on increasing in this regime. However, the rate of vapour removal remains greater than the rate of vapour generation and hence, no vapour gets accumulated within the porous bed. 

\begin{figure*}[!ht]
\centering
	\includegraphics[scale=0.45]{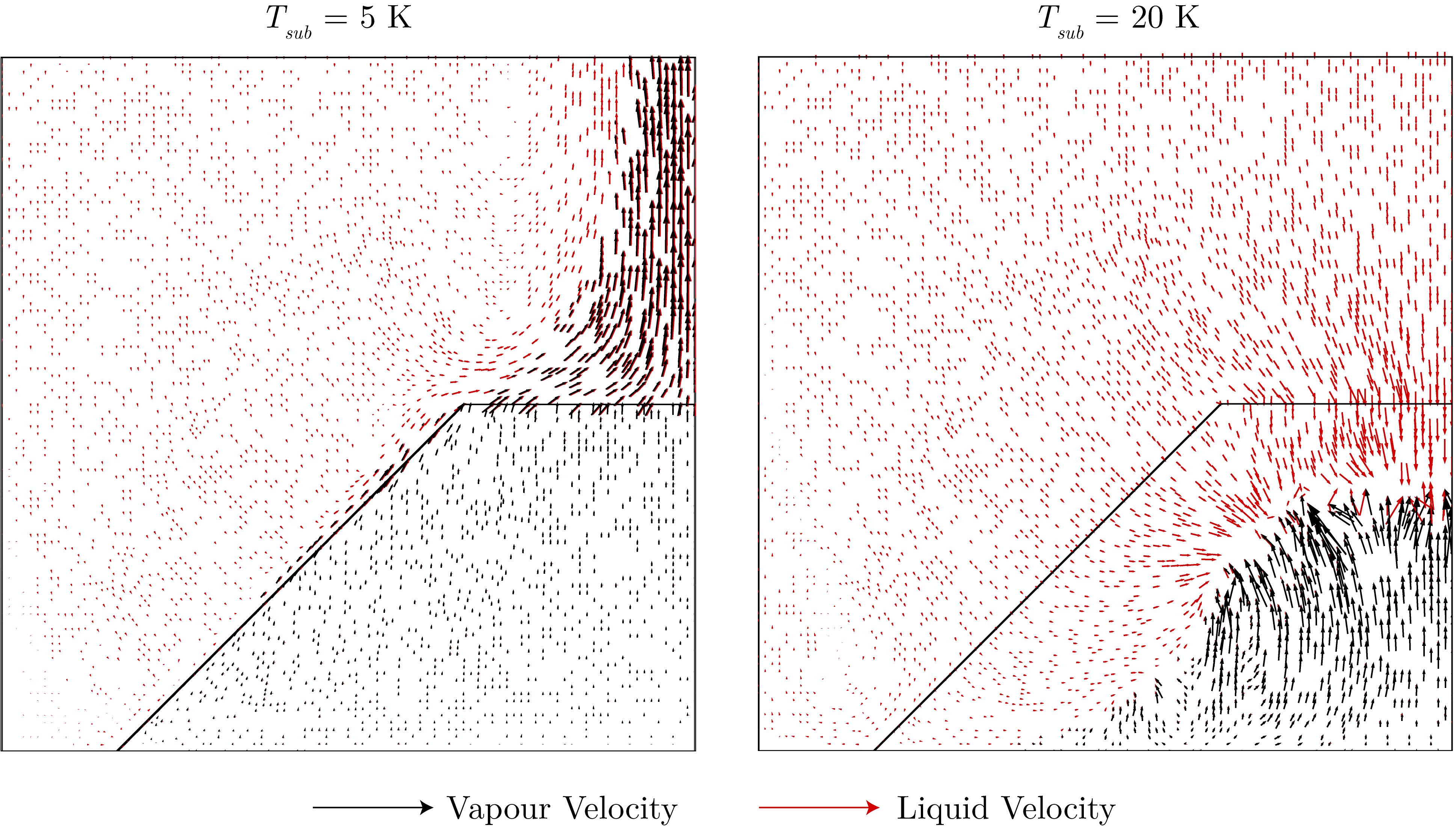}
	\caption{Liquid and vapour velocity vectors within the domain at dryout condition for two different coolant subcooling. Liquid-vapour counter-current flow restricts penetration of liquid coolant into and vapour removal from the porous bed. This causes vapour accumulation inside the porous bed leading to dryout. Shifting of the dryout zone towards the interior of the porous bed with increase in coolant subcooling is evident from this comparison.}
	\label{f:velocity}
\end{figure*}

\begin{figure*}[!ht]
\centering
	\includegraphics[scale=0.4]{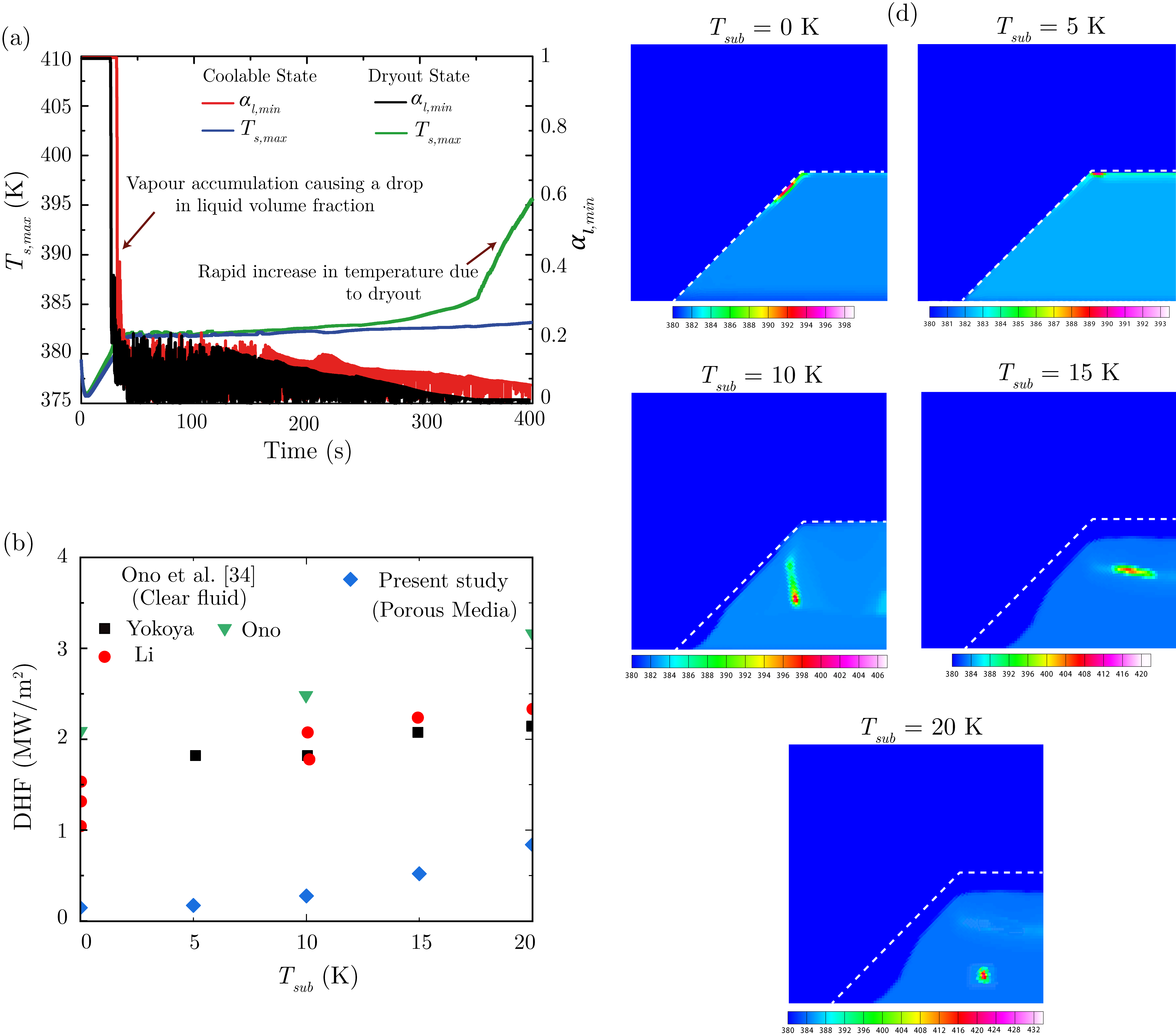}
	\caption{(a) Temporal change in $\alpha_{l,min}$ and $T_{s,max}$ at maximum coolable and minimum dryout power densities for $T_{sub}=10$K (b) Change in dryout heat flux with liquid subcooling as obtained in the present study in comparison to that observed experimentally in a clear fluid medium (redrawn with permission from Ono et al. \cite{ono2007}). (c) Shift in dryout zone within the porous bed with change in coolant subcooling. Temperature hotspots indicate the dryout zone. It is evident from these figures that the dryout power density increases progressively and the dryout zone shifts towards the interior of the porous bed with increase in coolant subcooling. Similarity to the effects of subcooling on dryout in clear fluids is also self-evident.}
	\label{f:dryout}
\end{figure*}

Figure \ref{f:velocity} shows the flow path of the liquid and vapour phases within the domain when subcooled coolant is considered. A constant inflow of subcooled coolant from the top boundary of the cavity results in the subcooled coolant coming in direct contact with the vapour formed within the porous bed. This results in condensation of the vapour due to interfacial heat transfer between the subcooled coolant and the vapour. The vapour generation and condensation phenomena, as such, takes place simultaneously within the porous bed such that the consequent transport processes are governed by the relative dominance of these phenomena. This is indicated by the rapid fluctuations in the magnitude of $\alpha_{l,min}$ for the subcooled coolants (Fig. \ref{f:dynamics}a). The magnitude of $\alpha_{l,min}$ drops to a very low value and continues to fluctuate for coolant subcooling up to 15 K. A significantly different characteristic is, however, evident for $T_{sub} = 20$ K. It is observed that $\alpha_{l,min}$ starts increasing after approximately 300 s, in contrast to the lower subcoolings, suggesting an enhanced impact of condensation. This rise in $\alpha_{l,min}$  continues till approximately 400 s after which it starts reducing again due to increased vapour generation. Liquid saturation distribution for $T_{sub} = 20$ K at different time instances corroborate this inference (see Fig. \ref{f:dynamics}a).

The temporal change of $T_{s,max}$ and $\alpha_{l,min}$ in Fig. \ref{f:dynamics} provides further insights into the effects of coolant subcooling on the heat removal process. The thermal equilibrium between the solid phase and the liquid coolant is observed to be attained at successively lower solid temperatures as the coolant subcooling is progressively increased. The steady-state temperature of the solid phase is also observed to progressively decrease as the coolant subcooling is increased. It is further observed that the time taken to reach a steady-state temperature becomes larger as the subcooling becomes higher. As such, the time interval till the onset of nucleate boiling in the porous bed - in case the power density is sufficient to cause boiling - also becomes longer with increase in subcooling. This is also evident from the drop in the magnitude of $\alpha_{l,min}$ in Fig. \ref{f:dynamics}a. The magnitudes of $\alpha_{l,min}$ further suggest a relatively lower vapour presence in the porous bed at higher coolant subcoolings. As a result, the increase in steady-state solid temperature is observed to be lower at larger coolant subcoolings for a constant power density, even when boiling takes place. This is also evident from the temperature rise curve in Fig. \ref{f:boilingcurve} and is qualitatively similar to the effects of subcooling on the boiling characteristics in a clear fluid medium.

It is, thus, inferred that larger amount of heat is removed from the porous bed as the coolant subcooling is increased. This leads to substantial quantitative differences between the temperature rise curves for different subcoolings, although the qualitative nature of the curves are similar. Similar effects of coolant subcooling on the boiling curve has been previously reported by various experimental studies \cite{el2016,ono2007}, although in different physical situations.

\subsection{Dryout assessment}
\label{sec:dryout}
The rate of vapour generation becomes large enough to exceed the vapour removal rate beyond a certain critical power density and in such a situation, vapour accumulation takes place within the porous bed. Heat removal reduces drastically in such vapour accumulated regions and as such, the temperature starts to increase rapidly whenever the heating power density exceeds this critical value. Such a situation is referred to as dryout of the bed and can be identified using Fig. \ref{f:boilingcurve} from the sharp change in the slope of the curve. Dryout indicates the maximum amount of heat that can possibly be removed from the bed in a given situation and any further increase in the power density would result in unhindered rise in the solid temperature and have potentially hazardous consequences (see Section \ref{sec:introduction}). The corresponding power density is termed as the dryout power density. 

The critical and dryout power densities for all coolant subcooling are identified following the methodology detailed in Section \ref{sec:dryout_identify}. A representative case is shown in Fig. \ref{f:dryout}a for coolant subcooling of 10 K. It can be observed that $\alpha_{l,min}$ starts reducing below 1.0 within a short period of time indicating vapour generation inside the bed. At the critical power density, $\alpha_{l,min}$ reduces to a very small magnitude and oscillates steadily about a mean value. The corresponding $T_{s,max}$ remains near the saturation temperature indicating removal of adequate thermal energy from the porous bed. At the dryout power density, however, $\alpha_{l,min}$ is reduces to zero suggesting vapour accumulation within the bed. The corresponding change in $T_{s,max}$ indicates a rapid rise indicative of dryout in the bed. 

It can be observed from Fig. \ref{f:boilingcurve} that the critical power density (or CHF) and hence, the dryout power density (or DHF) is significantly dependent on the coolant subcooling. Both quantities are observed to increase substantially as the subcooling is gradually increased. The change in DHF with coolant subcooling is plotted separately in Fig. \ref{f:dryout}b and compared with experimentally observed DHFs in case of clear fluids \cite{ono2007}. As can be observed, the increase in DHF with subcooling, as obtained from the present computational analysis with heat-generating porous media, has a qualitatively similar nature to the increase in DHF for clear fluids. However, the magnitude of DHF for porous media remains lower than that for clear fluids, similar to that observed in reported experimental studies \cite{fukusako}. 

The change in DHF (or dryout power density) with coolant subcooling is also accompanied by a shift of the dryout location in the porous bed. Vapour formed within the porous bed remains confined to inner regions of the bed as the subcooling is progressively increased (see Figs. \ref{f:dynamics}c and \ref{f:velocity}). Vapour accumulation necessary for causing dryout, therefore, also takes place in such regions. As a result, dryout is observed to take place at inner regions of the bed as the liquid coolant subcooling is progressively increased. This is represented in Fig. \ref{f:dryout}d using distribution of solid phase temperature  at dryout conditions for different coolant subcooling. Temperature hotspots indicate the region where dryout has taken place.

\section{Conclusions}
\label{sec:conclusions}
The present analysis fundamentally analyses the effect of liquid coolant subcooling on boiling heat transfer and multiphase flow, as well as  dryout in porous media with internal heat generation. A detailed computational study is carried out considering a truncated conical shaped, heat-generating porous bed submerged in a fluid-filled cylindrical cavity using a generalised computational model implemented in ANSYS FLUENT. 

It is observed from the analysis that the qualitative nature of boiling heat transfer characteristics remain similar, irrespective of the subcooled state of the coolant. Quantitatively, however, it is observed that coolant subcooling significantly influence the boiling heat transfer characteristics. Coolant subcooling is also observed to influence the heat transfer dynamics. Greater heat transfer occurs from the solid phase of the porous bed to the liquid coolant when the latter is in a subcooled state. A larger heating power density is, thus, necessary for the onset of boiling in case of subcooled coolants. Subcooled state of the coolant also enhances condensation of the vapour generated which, in turn, allows removal of larger amount of heat from the porous bed.  

The combined impact of these effects cause the dryout power density (or DHF) to increase substantially with coolant subcooling. The dryout power density increases from 2.54 MW/m\textsuperscript{3} for saturated water to 2.93 MW/m\textsuperscript{3} at 5 K subcooling and thereafter, to 14.36 MW/m\textsuperscript{3} for 20 K subcooling. It is also observed that the dryout zone develops in the inner regions of the bed with increase in coolant subcooling.

\vspace{-10pt}
\bibliographystyle{elsarticle-num-names}
\bibliography{references.bib}

\end{document}